# Simplified Polarization-Encoding for BB84 QKD Sourced by Incoherent Light of a Silicon Emitter

Florian Honz[(1)], Nemanja Vokić[(1)], Philip Walther[(2)], Hannes Hübel[(1)], and Bernhard Schrenk[(1)]

[(1)] AIT Austrian Institute of Technology, 1210 Vienna, Austria. Author e-mail: florian.honz@ait.ac.at
[(2)] University of Vienna, Faculty of Physics, 1090 Vienna, Austria.

**Abstract** *We investigate a polarization-encoded BB84-QKD transmitter that is simplified from an architectural and technological point-of-view, demonstrating a silicon emitter sourcing a low-complexity polarization modulator for secure-key generation at a raw-key rate of 2.8kb/s and QBER of 10.47%, underpinning the feasibility of an all-silicon QKD transmitter.*                                                                                 ©2023 The Author(s)

**Introduction**
With quantum computating reaching new record-breaking numbers of qubits nearly every month [1], it is clear that quantum computers are on the brink of first real-world deployments. Despite these rapid advances, the encryption standards we are currently using to secure our communications as well as our personal data are in jeopardy. Quantum key distribution (QKD) is on the forefront of becoming the new encryption standard, by virtue of offering provable security and being well tested – with some QKD systems already being operated successfully for years.

One of the most attractive targets for eavesdroppers are datacenters, simply because of the vast amount of data which can be tapped and possibly decrypted at a later time. Due to this reason, zero-trust models start to inundate this realm. Introducing QKD to this highly specialized environment aggregating 10,000 servers or more [2] remains highly challenging. Cost-effective deployment would require the QKD systems to undergo a disruptive down-scaling of their footprint while withstanding elevated temperatures and being robust to increased EMI. Ideally, QKD hardware is compatible with modern packaging trends such as co-packaged optics [3], which require a chiplet QKD approach.

In this work, we investigate a simplified QKD transmitter that addresses these needs. We show low-complexity polarization-encoding in a BB84 QKD transmitter sourced by a silicon light emitter. We show that a raw-key rate of 29.3 kb/s with a QBER of 10.63% can be in principle achieved for a 2-nm wide incoherent light source over 256 m of fiber, or 2.8 kb/s at a QBER of 10.47% using a silicon source based on a Ge-on-Si PIN junction.

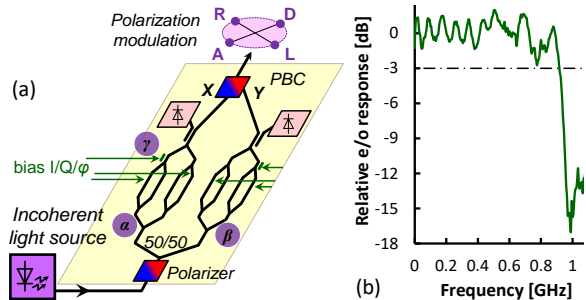

**Fig. 1:** (a) Polarization encoder for BB84 QKD and (b) its electro-optic response.

**Intra-Datacenter QKD with Ge-on-Si Source**
Photonic integration is considered a key enabler for the miniaturization of QKD systems. Various research efforts have focused on the shoehorning of QKD transmitters or receivers on PICs, leading to impressive demonstrations for both, DV- and CV-QKD [4-7]. However, as for classical optical telecommunication systems, none of these can truly provide a monolithically integrated solution that is able to accommodate the required electronic circuitry – a necessity to avoid exposure of critical interfaces to any empowered eavesdropper.

Toward this direction, we recently proposed a monolithic silicon QKD transmitter, which not only caters for quantum state preparation, but also for light generation. Besides, silicon is known for its compatibility with electronic co-integration. In particular, we proved a Ge-on-Si PIN junction to be a suitable light source for QKD building on independent polarization modulation using four Mach-Zehnder modulators (MZM) in combination with passive polarizing optics [8]. Here, we will simplify the modulation scheme towards interferometric polarization modulation while investigating the performance limitations in combination with an incoherent light source based on amplified spontaneous emission (ASE) and the Ge-on-Si light source, respectively.

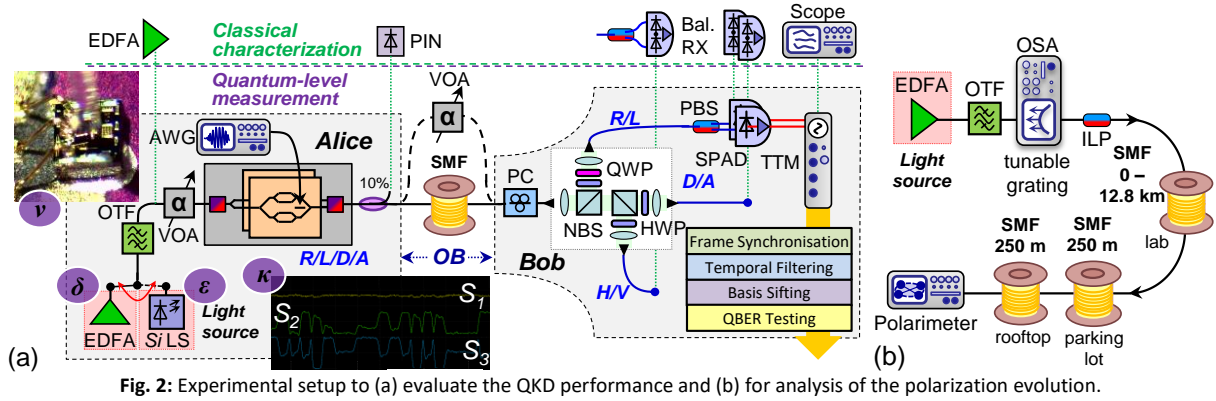

**Fig. 2:** Experimental setup to (a) evaluate the QKD performance and (b) for analysis of the polarization evolution.

**Simplified Polarization Modulation**

The BB84 protocol based on polarization encoding necessitates four basis states in two non-orthogonal polarization bases, as provided through the polarization states **R**, **L**, **A** and **D** in the circular and diagonal bases. For this purpose, we employ a dual-polarization I/Q modulator, as specified in the OIF implementation agreement [9]. Through its integrated nested I/Q modulators ($\alpha$, $\beta$ in Fig. 1a), the **X** and **Y** polarizations can be independently modulated. Since the phase $\varphi$ between the I and Q signals at the child-MZMs can be adjusted, we used this degree of freedom to generate the desired polarization states in the **X / Y** plane. We thereby apply only the DC biases for the I and Q electrodes, in order to balance the power levels in the **X** and **Y** polarizations. Instead of I and Q, we modulate the phase $\varphi$ ($\gamma$ in Fig. 1a). In this way we rotate the polarization state along the $S_2/S_3$ plane in the Stokes space, effectively creating a four-state protocol with **A** (-45°), **D** (+45°), **R** (right-circular) and **L** (left-circular). As it is reported in Fig. 1b, the electro-optic response of the phase section has a sufficiently wide bandwidth of 920 MHz to allow for modulation up to 1 Gb/s.

**Fiber Depolarization Effects**

As a result of the already mentioned broadband and incoherent nature of the Ge-on-Si light source, special attention needs to be paid to the propagation of the QKD (polarization) state along the transmission fiber. The different group velocities of the two principal polarization modes of a standard single mode fiber (SMF) lead to polarization mode dispersion (PMD), while the coherence time of the broadband light source is also exceeded by the differential group delay, leading to depolarization [10, 11].

To characterize the depolarization effect, we used an L-band EDFA, whose incoherent ASE is spectrally filtered and transmitted over a fiber link before being acquired by a polarimeter (Fig. 2b). The spectral filtering was performed by first employing a bandwidth-tunable optical filter (OTF) with a bandwidth of $\Delta\Lambda$ = 16 nm. We then used the tunable grating of an optical spectrum analyzer (OSA) for a wavelength-swept slicing of the filtered ASE spectrum into narrower slices, having a bandwidth of $\Delta\lambda$ = 1 nm. The sweep was performed from $\Lambda_{min}$ = 1569 nm to $\Lambda_{max}$ = 1585 nm and repeated every 5 minutes. We polarized the narrow slices (ILP) before transmission over SMF spools with different lengths of up to 12.8 km. Two 250-m long deployed fiber spans over a rooftop and a parking lot served a faster polarization drift evolution due to their exposure to vibrations, strain and temperature variations.

The wavelength-dependent evolution of the polarization state is reported in Fig. 3a for a measurement over 5 hours. As expected, the polarization states separate, which is associated to depolarization, and the long-term drift increases as the fiber length does. This indicates that a transmitted quantum state will lose its degree of polarization, leading to an increase in QBER due to a reduced polarization extinction when analyzing the received BB84 state in the two non-orthogonal polarization bases.

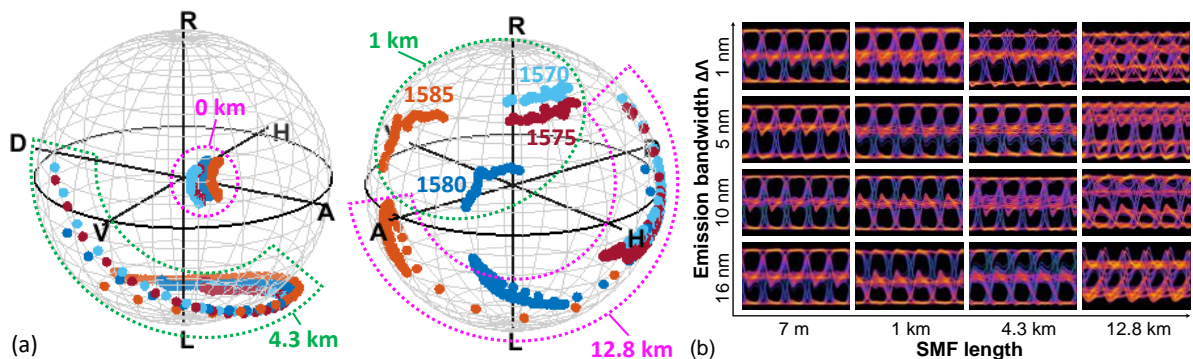

**Fig. 3:** (a) Evolution of the polarization states for various wavelengths and SMF lengths from 0 - 12.8 km. (b) Eye diagrams showing the depolarization for different SMF lengths and bandwidths under polarization encoding at 100 MHz symbol rate.

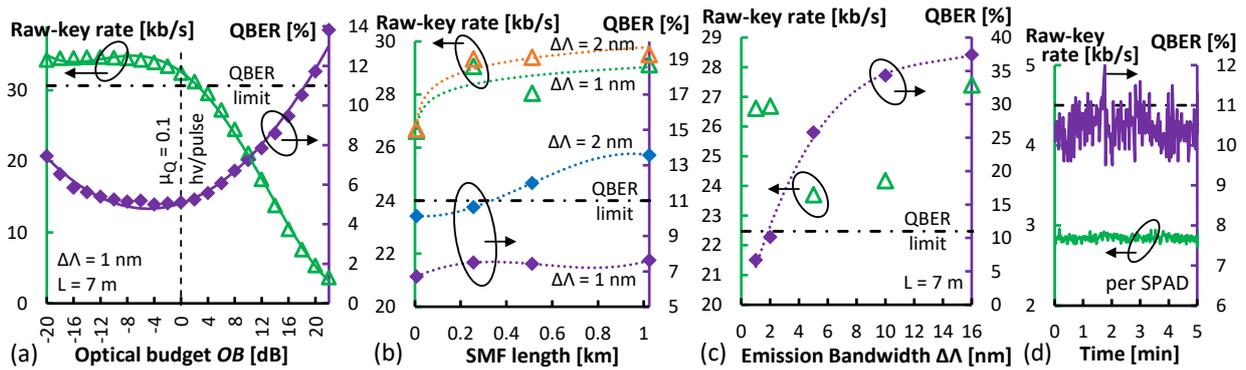

**Fig. 4:** (a) Raw-key rate and QBER for the ASE-sourced QKD transmitter at 1 GHz symbol rate and dependence on (b) the fiber length and (c) optical emission bandwidth ΔΛ. (d) QBER and raw-key rate for the Ge-on-Si sourced QKD transmitter.

We further investigated the depolarization effect in the optical eye diagram of the transmitted BB84 signals (Fig. 3b), using the identical setup as for the QKD evaluation (Fig. 2a) under classical signal launch conditions. As can be seen, the signal integrity decreases with longer fiber reach and wider optical bandwidth of the incoherent light source, resulting in a migration from a clearly open 3-level eye diagram to a degenerated one with only two levels.

These findings drive the conclusion that narrow emission bandwidths for the light sources are paramount for obtaining a good QKD performance. However, narrowing the emission spectrum through optical filtering comes with a reduction in emission power. We therefore conducted a prepare-and-measure QKD performance evaluation for an incoherently sourced BB84 transmitter employing polarization encoded quantum states (Fig. 2a).

**Performance of Incoherently Sourced QKD**

Our QKD transmitter builds on incoherent light seeding the simplified I/Q based polarization modulator in Fig. 1a. We used either the ASE of an L-band EDFA ($\delta$) or the Ge-on-Si PIN emitter ($\varepsilon, \nu$) as the light source, followed by an OTF to set ΔΛ. A variable optical attenuator (VOA) ensures transmission of the quantum signal at a power level of $\mu_Q$ = 0.1 photons/symbol at Alice' output. We then generate the polarization states in the $S_2/S_3$ ($\kappa$) plane via phase modulation in the I/Q modulator. The fiber length and the optical budget of the link have been varied for the purpose of performance assessment.

On Bob's side, we first align the state of the input polarization to the axes of our polarization analyzer through manual polarization control (PC). The BB84 state analyzer then transmits the signals according to their polarization state to one of two free-running InGaAs SPADs (10% efficiency, 25 μs dead time, 500-600 dark counts/s). The detection events were recorded by a time-tagging module (TTM) and a real-time QBER evaluation was performed after temporal filtering within 50% of the symbol period.

At first, we evaluated the compatible optical link budget (OB) between Alice and Bob for a back-to-back scenario without transmission fiber, employing the ASE source and a transmitter symbol rate of 1 GHz (Fig. 4a). As can be seen, we can accommodate up to 18.5 dB of OB before reaching the QBER limit of 11% at which a secret key can still be generated [12]. At negative optical budgets, meaning $\mu > \mu_Q$, an increase in QBER (◆) can be noticed due to detector saturation, together with a saturation of the raw-key rate (△).

Next, we used the same parameters to determine the dependence of the QBER on the SMF length and emission bandwidth ΔΛ (Fig. 4b), confirming a QBER increase with transmission reach and filter bandwidth, as expected from Fig. 3b. For ΔΛ = 2 nm we can generate a secret key for lengths up to 256 m, while we can go up to
1 km for ΔΛ = 1 nm. We found a steep increase in QBER as a function of the emission bandwidth ΔΛ, reaching 25% for ΔΛ = 5 nm (Fig. 4c).

Finally, we employed the proposed Ge-on-Si PIN emitter as the light source. When forward biased, this silicon source shows a LED-like behaviour, emitting light in the C+L bands. The peak emission wavelength was 1581 nm and the maximum output power is -70 dBm at a forward current of 46 mA. Due to the necessary tight filtering of the optical bandwidth to ΔΛ = 2 nm, we were forced to switch to a transmit symbol rate of 100 MHz, still being restricted in $\mu$ to less than $\mu_Q$ due to the limited output power of the emitter. Nonetheless, we were able to obtain key generation with an average QBER of 10.47% (3σ = 1.41%) at a raw-key rate of 2.8 kb/s (Fig. 4d).

## Conclusion

We successfully employed a simplified polarization modulator for state preparation in a BB84 QKD transmitter together with a silicon light source. We have investigated the restrictions that apply to incoherent light sources for such a QKD approach, rendering intra-datacenter scenarios feasible for bandwidths of up to 2 nm. We further demonstrated secure-key generation for the Ge-on-Si PIN emitter through a raw-key rate of 2.8 kb/s at a QBER of 10.47%, proving the feasibility of a potential all-silicon QKD transmitter.

## Acknowledgements

This work was supported by the European Research Council (ERC) under the EU Horizon-2020 programme (grant agreement No 804769) and by the Austrian FFG Research Promotion Agency and NextGeneration EU (grant agreement No FO999896209).